\documentclass[aps,pra,nobibnotes,twocolumn,superscriptaddress]{revtex4-1}
\usepackage{graphicx}
\usepackage{mathrsfs}
\usepackage{bm}
\usepackage{amsmath}
\usepackage{dcolumn}
\usepackage{epstopdf}
\usepackage{dsfont}
\usepackage{amssymb}
\usepackage{tabularx}
\usepackage{array}
\usepackage{float}
\usepackage{color}
\usepackage{epstopdf}
\usepackage{mathrsfs}
\usepackage[colorlinks, linkcolor=blue,anchorcolor=blue,citecolor=blue,urlcolor=blue]{hyperref}

\usepackage{xcolor}

\begin{document}

\preprint{APS/123-QED}

\title{Quantum phase transitions of interacting bosons on hyperbolic lattices}

\author{Xingchuan Zhu}
\affiliation{Department of Physics, Beijing Normal University, Beijing, 100875, China}

\author{Jiaojiao Guo}
\affiliation{School of Physics, Beihang University,
Beijing, 100191, China}

\author{Nikolas P. Breuckmann}
\affiliation{Department of Physics and Astronomy, University College London, WC1E 6BT London, United Kingdom}

\author{Huaiming Guo}
\email{hmguo@buaa.edu.cn}
\affiliation{School of Physics, Beihang University,
Beijing, 100191, China}

\author{Shiping Feng}
\affiliation{Department of Physics, Beijing Normal University, Beijing, 100875, China}

\pacs{ 03.65.Vf, 
 67.85.Hj 
 73.21.Cd 
 }

\begin{abstract}
The effect of many-body interaction in curved space is studied based on the extended Bose--Hubbard model on hyperbolic lattices. Using the mean-field approximation and quantum Monte Carlo simulation, the phase diagram is explicitly mapped out, which contains the superfluid, supersolid and insulator phases at various fillings. Particularly, it is revealed that the sizes of the Mott lobes shrink and the supersolid is stabilized at smaller nearest-neighbor interaction as $q$ in the Schl\"afli symbol increases. The underlying physical mechanism is attributed to the increase of the coordination number, and hence the kinetic energy and the nearest-neighbor interaction. The results suggest that the hyperbolic lattices may be a unique platform to study the effect of the coordination number on quantum phase transitions, which may be relevant to the experiments of ultracold atoms in optical lattices.
\end{abstract}

\maketitle

\section{Introduction}

Quantum gravity is an exciting area to combine both quantum field theory and general relativity~\cite{1983Quantum,Rovelli2008}. Due to the incompatibility of the two theoretical frameworks, constructing a unified theory remains elusive. Remarkable progress in quantum simulations has allowed to realize curved space on table-top experimental setups~\cite{hu2019quantum,franz2018}, which opens the door to explore novel quantum phenomena beyond flat spaces. As quantum many-body physics is a main theme in condensed matter physics~\cite{Quintanilla_2009}, an interesting question is how interacting particles behave in non-Euclidean spaces.

While the surface of a sphere has positive Gaussian curvature, a surface in hyperbolic space with constant negative curvature can not be realized in Euclidean space without distortion~\cite{ratcliffe1994,reynolds1993hyperbolic,wilson_2007}.
There are only three regular tilings of Euclidean space (square, triangular, hexagonal) but infinitely many regular tilings of hyperbolic space.
The hyperbolic lattices have the remarkable property that a compactified manifold has genus $g>1$ and a comparable number of sites reside on the boundary of an open hyperbolic lattice, generating strong boundary effect. It is highly expected that the physical properties of strongly correlated systems on hyperbolic lattices can be drastically different from their flat-space counterparts.

Indeed significant efforts have been devoted to the studies of statistical models on hyperbolic lattices, such as ferromagnetic Ising model~\cite{Rietman_1992,Nikolas_2020,Nishino_2010,rietman1992,shima2006}, $XY$ model~\cite{PhysRevE.79.060106}, percolation, diffusion~\cite{PhysRevE.77.022104}, clock model~\cite{PhysRevE.77.041123,PhysRevE.80.011133} et al.. The ferromagnetic Ising model on hyperbolic planes has been investigated thoroughly, and all works reveal the phase transition follows a mean-field behavior, i.e., the critical exponents and critical temperatures obtained are close to the mean-field ones. Specially self-dual hyperbolic lattices are different from the flat-space counterpart, where two distinct critical temperatures $T_c$ and $\overline{T}_c$ exist, related to one another by the Kramers--Wannier duality relation $\sinh(2J/T_c)\sinh(2J/\overline{T}_c)=1$~\cite{wu1996}. A new phase appears between $T_c$ and $\overline{T}_c$, which breaks translational symmetry, and consists of infinite many and large clusters of magnetized spins. Its existence is purely due to the negative curvature of the embedding space, and has been proved for a hyperbolic plane with free boundary condition. However obstructed by the difficulty to find large enough sizes, signatures of this intermediate phase is still lack on a compactified hyperbolic plane~\cite{Nikolas_2020,sausset2007}. Similarly, percolation on self-dual hyperbolic lattices also shows two distinct transitions~\cite{Baek_2009}.

Recent progress in circuit quantum electrodynamics has made the realization of hyperbolic lattices possible, where unusual gapped flat band for free itinerant electrons on hyperbolic analogues of the kagome lattice were discovered~\cite{kollar2019hyperbolic}. Several subsequent theoretical studies were motivated by this experimental breakthrough. Bloch band theory is generalized to hyperbolic lattices based on ideas from Riemann surface theory and algebraic geometry~\cite{maciejko2020hyperbolic}. Topological states of matter in hyperbolic lattices have been investigated by examining the topological protection of helical edge states and generalize Hofstadter's butterfly~\cite{PhysRevLett.125.053901}. In particular, using graph theory and differential geometry, quantum field theories in continuous negatively curved space has been formulated for quantum many-body systems on hyperbolic lattices~\cite{PhysRevA.102.032208}. Naturally, it is highly desirable to directly simulate the many-body models on hyperbolic lattices using exact numerical methods.

In this paper, we investigate interacting quantum particles on hyperbolic lattices based on the fundamental Bose--Hubbard model, which was first derived to describe ultracold bosons in optical lattices~\cite{2012Quantum,2005Quantum,PhysRevLett.81.3108,PhysRevA.63.053601}. We employ the mean field approximation, the second-order perturbation theory and quantum Monte Carlo (QMC) simulations to study the extended Bose--Hubbard model. The role of the coordination number on the quantum phase transitions is specially investigated. We reveal that the sizes of the Mott lobes shrink and the supersolid is stabilized at smaller nearest-neighbor (NN) interaction as $q$ in the Schl\"afli symbol (see Section \ref{sec:hyperbolic_latices}) increases. These behaviors can be well understood in terms of the increase of the kinetic energy and the NN interaction, which are proportional to the coordination number.
Our results are closely relevant to the experiments of
ultracold atoms in optical lattices.

\section{Hyperbolic lattices}\label{sec:hyperbolic_latices}
The hyperbolic plane is a two-dimensional, homogeneous space that has a constant negative curvature.
It is distinguished from the Euclidean plane and spherical geometry which have zero curvature and constant positive curvature, respectively.
When embedded into a higher-dimensional Euclidean space, every point of the hyperbolic plane locally looks like a saddle point.
Due to the curvature, the hyperbolic plane can not be realized in Euclidean space without distortion.

A widely-used model of the infinite hyperbolic plane is the Poincar\'e disk, where the hyperbolic plane is mapped to the interior of a unit disk.
Hyperbolic geodesics are mapped onto circular arcs  that meet the bounding circle at right angles.
Hyperbolic circles are mapped onto Euclidean circles in the Poincar\'e disk model.

A hyperbolic surface can be tessellated by regular polygons placed edge-to-edge.
Each regular tiling can be labeled by the number of sides of the polygons~$p$ and the number~$q$ of polygons meeting at each vertex of the tiling.
This label is known as the Schl\"afli symbol $\{p,q\}$.
Unlike in Euclidean space, the sum of the angles of a triangle on a surface of negative curvature will be less than~$\pi$.
Hence, the numbers~$p$ and~$q$ satisfy the following relation,
\begin{eqnarray}\label{eq1}
\frac{2\pi}{p}+\frac{2\pi}{q}<\pi \Leftrightarrow (p-2)(q-2)>4.
\end{eqnarray}
It turns out that this equation is the only condition on~$p$ and~$q$, so that there are an infinite number of regular hyperbolic tilings.

\begin{figure}[htbp]
\centering
\includegraphics[width=9cm]{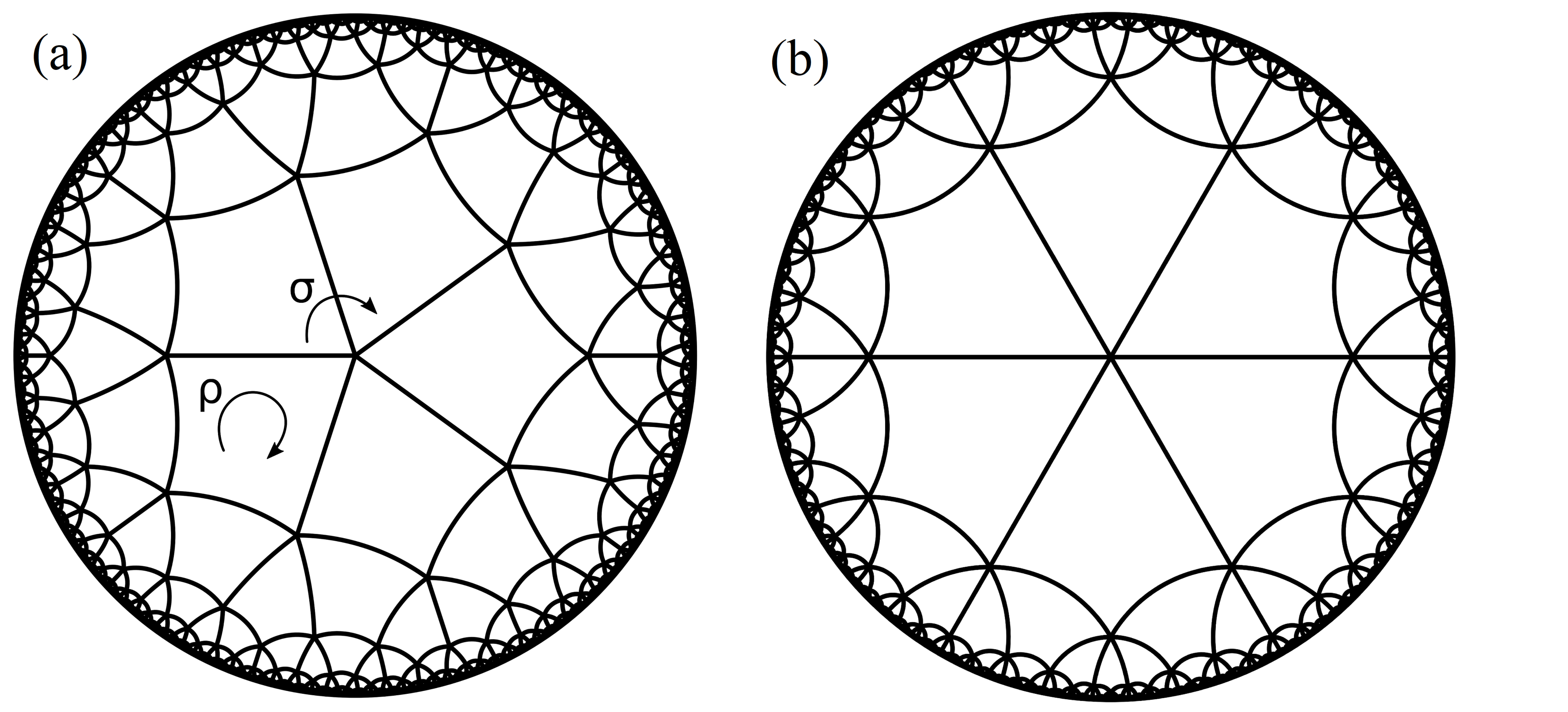}
\caption{The regular $\{4,5\}$-lattice (left) and the regular $\{4,6\}$-lattice (right) shown in the Poincar\'e disk model.
The group~$G_{p,q}$ of orientation-preserving symmetries are generated by the face-rotation~$\rho$ and vertex-rotation~$\sigma$.}
\label{fig1}
\end{figure}

A difficulty of hyperbolic space is that boundary effects are severe.
In $D$-dimensional Euclidean space a ball of radius $r$ has volume $\propto r^D$ and boundary of size $\propto r^{D-1}$ so that boundaries can be neglected in the thermodynamic limit.
This does not hold in hyperbolic space where the ratio between the size of the bulk and the size of the boundary is a constant.
In fact this constant can be larger than 1/2 so that such a model is dominated by the boundary.
In order to perform finite-size scaling it is therefore necessary to introduce periodic boundary conditions.
This confronts us with another problem, namely that translations in curved spaces do not commute.
This problem can be solved algebraically by considering the group of (orientation-preserving) symmetries of the lattice~\cite{magnus1974noneuclidean,Sausset_2007,Breuckmann2016,breuckmann2018phd}.
This is a triangle group which depends on the Schl\"afli symbol and which can be expressed as a finitely presented group as
\begin{align*}
G_{p,q} = \langle \rho, \sigma \mid \rho^p = \sigma^q = (\rho \sigma)^2 = e \rangle
\end{align*}
where $e$ denotes the neutral element of $G_{p,q}$.
The generator $\rho$ corresponds to a rotation around the center of a face and $\sigma$ corresponds to a rotation around an adjacent vertex (see Fig.~\ref{fig1}).
The vertices of the lattice are naturally identified with cosets of the subgroup~$\langle \sigma \rangle$ generated by~$\sigma$.
In order to introduce periodic boundaries we consider a normal subgroup $N$ of $G_{p,q}$ which only contains hyperbolic translations and no rotations.
The quotient group $G_{p,q}/N$ is then the symmetry group of a hyperbolic surface in which all points differing by an element in $N$ are identified.

\section{The extended Bose--Hubbard model and the QMC method}
We consider the interacting bosons on hyperbolic lattices in the grand canonical ensemble, whose basic physics is described by the following extended Bose--Hubbard model~\cite{PhysRevB.75.214509,2007honeycomb,2005Supersolid,PhysRevB.84.174515,PhysRevLett.97.087209,PhysRevLett.94.207202}:
\begin{align}\label{eq2}
H=&-t\sum_{\langle i,j\rangle} (b_i^{\dagger}b_{j}+h.c.) +\frac{U}{2}\sum_i n_i(n_i-1)  \\ \nonumber
&+\sum_{\langle i,j\rangle} V n_in_j-\mu \sum_i n_i,
\end{align}
Here $b_i$ ($b_i^{\dagger}$) is the bosonic annihilation (creation) operator on site $i$. These operators obey the commutation relations $[b_i,b_j^{\dagger}]=\delta_{ij}$. $n_i=b_i^{\dagger}b_i$ is the number operator of bosons. $\langle i,j\rangle$ runs over all NN pairs. The first term in Eq. (\ref{eq2}) corresponds to the NN hopping of bosons, with
amplitude $t$, which we taken as the unit of energy $t = 1$.
The second term in Eq. (\ref{eq2}) represents the on-site interaction with strength $U$.
The next line of the Hamiltonian describe the NN interaction and the on-site potential, with strength $V$ and the chemical potential $\mu$, respectively.

In the following discussions, we employ the approach of stochastic series expansion (SSE) quantum Monte Carlo (QMC) method~\cite{sandvik2002,syljuasen2003} with directed loop updates to study the model in Eq.~(\ref{eq2}). The SSE method expands the partition function in power series and the trace is written as a sum of diagonal matrix elements. The directed loop updates make the simulation very efficient~\cite{Bauer2011,fabien2005,pollet2004}. Our simulations are on finite lattices with periodic boundary condition. There are no approximations
causing systematic errors, and the discrete configuration space can be sampled without floating
point operations. The temperature is set to be low enough to obtain the ground-state properties. For such bosonic systems, the notorious sign problem in the QMC approach can be avoided.

\section{The Mean-field approximation}

\begin{figure}[htbp]
\centering \includegraphics[width=6.cm]{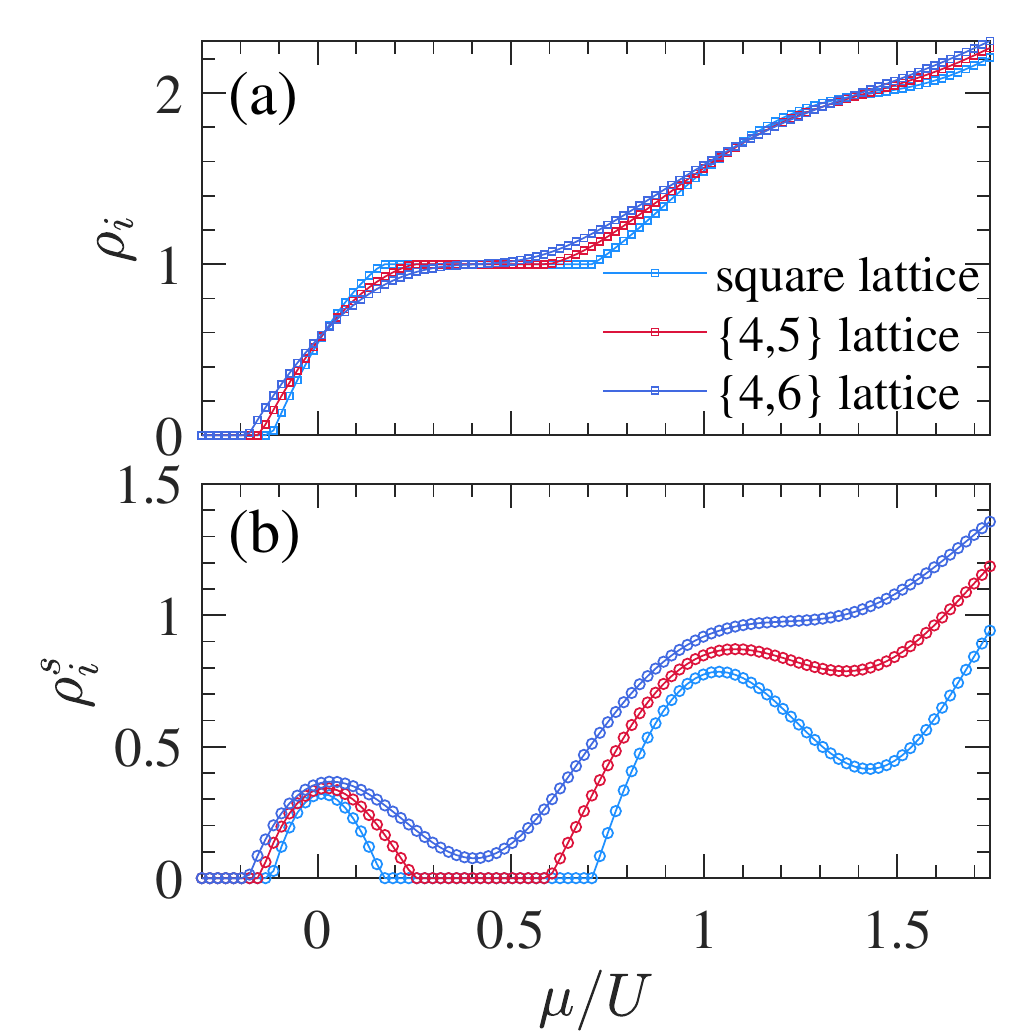} \caption{The mean-field average density $\rho_i$ (top) and superfluid density $\rho_i^s$ (bottom) as a function of $\mu/U$ at fixed $t/U = 0.03$.  }
\label{fig2}
\end{figure}

The product of two bosonic operators can be decoupled in the following mean-field channel~\cite{PhysRevB.40.546,Sheshadri_1993}:
\begin{align}\label{eq3}
b_i^{\dagger}b_j = \langle b_i^{\dagger} \rangle b_j+b_i^{\dagger}\langle b_j \rangle-\langle b_i^{\dagger} \rangle \langle b_j \rangle
\end{align}
The average value $\langle b_i^{\dagger} \rangle$ represents the superfluid order parameter $\Psi$ that characterizes the superfluid phase. It is zero in the insulating region of the phase diagram, and has a nonzero value in the superfluid state where the quantum fluctuation of the boson number is large. Moreover, $|\Psi|^2$ represents the local density of the bosons in the condensate state.

In the mean-field approximation, the Bose--Hubbard Hamiltonian described by Eq.~(\ref{eq2}) turns into a sum of the following single-site terms:
\begin{align}\label{eq4}
H_i^{MF}=-zt(\Psi^* b_i^{\dagger}+\Psi b_i -|\Psi|^2)+H_i^{loc},
\end{align}
where $z$ represents the number of NN sites, and are $4$, $5$ and $6$ for $\{4,4\}$, $\{4,5\}$ and $\{4,6\}$ hyperbolic lattices, respectively. Since $H_i^{loc}=\frac{U}{2}(n_i-1)n_i-\mu n_i$ is diagonal in the basis $\{|N_i\rangle\}$, we have
\begin{align}\label{eq5}
\langle N_i | H_i^{loc} |N_i\rangle=\frac{U}{2}(N_i-1)N_i-\mu N_i.
\end{align}
Hence the matrix elements of the mean-field Hamiltonian $H_i^{MF}$ in the occupation number basis $\{|N_i\rangle\}$ are as follows:
\begin{align}\label{eq6}
&\langle N_i |H_i^{MF} |N_i\rangle=\frac{U}{2}(N_i-1)N_i-\mu N_i+zt|\Psi|^2, \nonumber\\
&\langle N_i+1 |H_i^{MF} |N_i\rangle=-zt\Psi^* \sqrt{N_i+1}, \\
&\langle N_i-1 |H_i^{MF} |N_i\rangle=-zt\Psi \sqrt{N_i}, \nonumber
\end{align}
and all other ones are zero. For bosons, the occupation number $N_i$ on each site varies from $0$ to $\infty$. We diagonalize the Hamiltonian Eq.~(\ref{eq6}) in a truncated basis $|N_i\rangle$ with $N_i=0,1,\cdots,N_i^{max}$, and the ground state of the mean-field Hamiltonian writes as,
\begin{align}\label{eq7}
|G^i\rangle = \sum_{N_i=0}^{N_i^{max}} \alpha_{N_i} |N_i\rangle
\end{align}
with $\alpha_{N_i}$ the coefficients of the lowest eigenvalue of the Hamiltonian matrix.
Then the order parameter $\Psi$ in the ground state is,
\begin{align}\label{eq8}
\Psi=\langle G^i |b_i^{\dagger}|G^i\rangle=\sum_{N_i=0}^{N_i^{max}-1} \alpha_{N_i}\alpha_{N_i+1}^*\sqrt{N_i+1}.
\end{align}
By combining the Hamiltonian matrix in Eq.~(\ref{eq6}) and the formula for the order parameter in Eq.~(\ref{eq8}), $\Psi$ can be determined self-consistently. With the coefficients $\alpha_{N_i}$ and the order parameter $\Psi$, the average density
\begin{align}\label{eq9}
\rho_i=\langle n_i \rangle=\sum_{N_i=1}^{N_i^{max}} |\alpha_{N_i}|^2 N_i,
\end{align}
and the condensate component of the superfluid density on the site $i$
\begin{align}\label{eq10}
\rho_i^s=|\Psi|^2,
\end{align}
are directly obtained.

Figure \ref{fig2} plots the mean-field average density $\rho_i$ and superfluid density $\rho_i^s$ as a
function of $\mu/U$ at fixed $t/U = 0.03$. $\rho_i$  exhibits a sequence of plateaus at integer fillings, on which $\rho_i^s$ vanishes.
The plateaus correspond to the incompressible Mott insulators. By collecting the positions of the plateaus at different $t/U$, the mean-field phase diagram in the $(t/U,\mu/U)$ plane is mapped out. As shown in Fig.\ref{fig3}, the phase diagram is composed of a sequence of Mott insulating lobes, whose sizes shrink as $q$ in the Schl\"afli symbol increases. The phase boundaries can also be analyzed using the second-order perturbation theory~\cite{1990Second,1992second}, and the results are almost the same with those from the mean-field theory.

\section{The QMC results}
\begin{figure}[htbp]
\centering \includegraphics[width=9.cm]{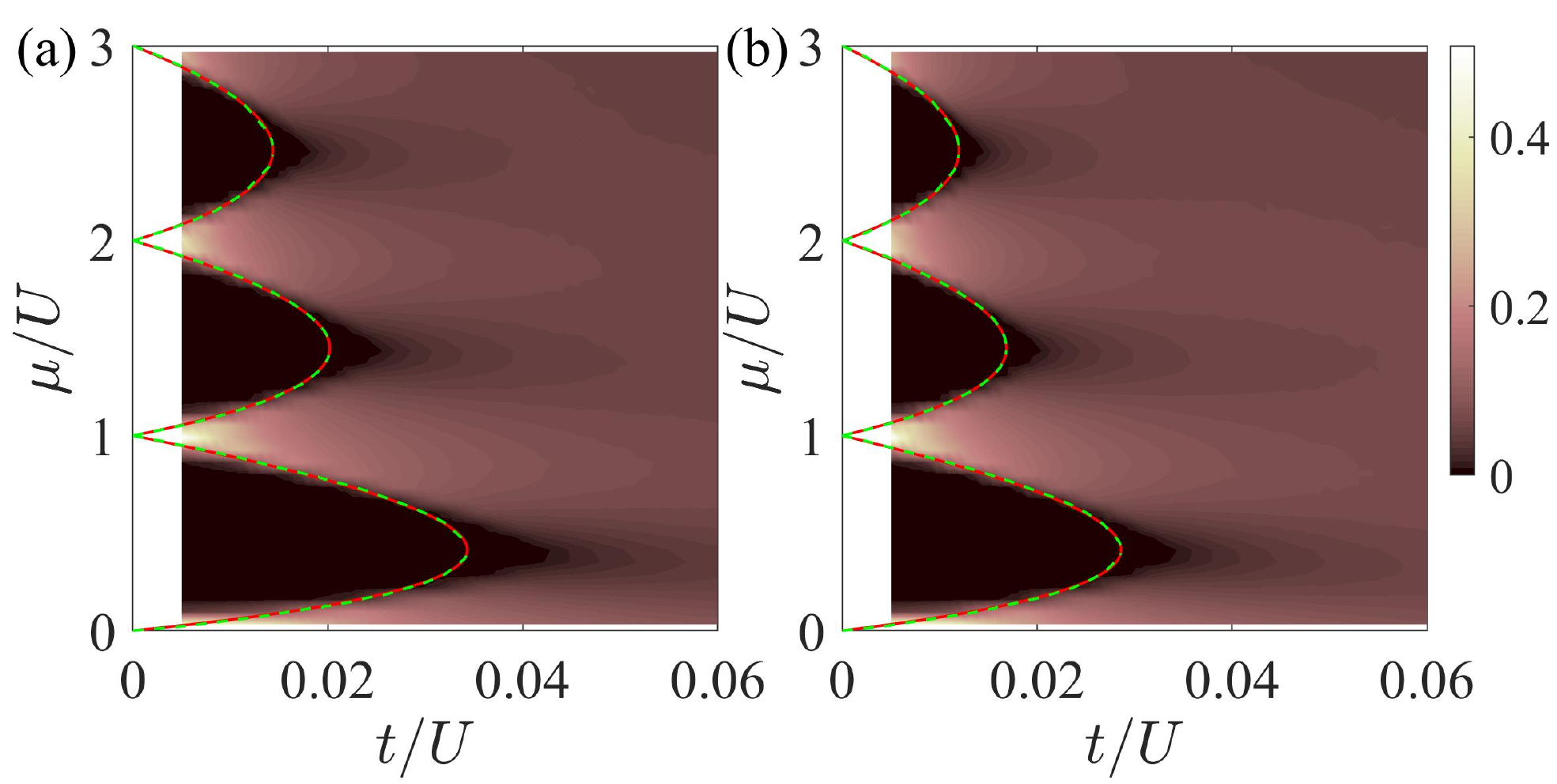} \caption{Phase diagram of the Bose--Hubbard model in the $(t/U, \mu/U)$ plane on the hyperbolic lattice with the Schl\"ali symbol: (a) \{4,5\} and (b) \{4,6\}. The false color represents the value of the compressibility $\kappa=\frac{d\rho}{d\mu}$. An insulator
is characterized by $\kappa=0$, while a superfluid phase by a finite $\kappa$. The red lines represent the results of the mean-field theory, and the green dotted lines are from the second-order perturbation theory. They are almost the same, and are indistinguishable in the figures.}
\label{fig3}
\end{figure}

We first consider the case with $V=0$. In the atomic limit $t=0$, whether a boson can be added to the $j$th site with $n_j$ bosons is determined by the energy difference $\Delta E=E(n_j+1)-E(n_j)=-\mu+Un_j$ with $E(n_j)=-\mu n_j+\frac{U}{2}n_j(n_j-1)$ the total energy of the bosons on the $j$th site. If the total energy is lowered, i.e., $\Delta E<0$, one more boson can be added to the site. Thus $\mu/U=n_j (n_j=0,1,2,...)$ separates different insulating phases at integer fillings. Next we turn on the hoppings and the phase diagrams obtained from QMC simulations for $\{4,5\}$ and $\{4,6\}$ hyperbolic lattices are shown in Fig.\ref{fig3}. It contains incommensurate superfluid and insulators at integer fillings. Although each insulator in the atomic limit persists, its range along the $\mu/U$ axis is reduced and incommensurate superfluid regions appear between the commensurate insulating regions. The phase boundary between Mott insulator and superfluid phase has a lobelike shape. As $q$ in the Schl\"afli symbol [here $q=5,6$ in Fig.1(a) and (b), respectively] increases, the sizes of the lobes shrink, and the critical hopping amplitude to break the Mott insulator decreases.

\begin{figure}[htbp]
\centering \includegraphics[width=6.5cm]{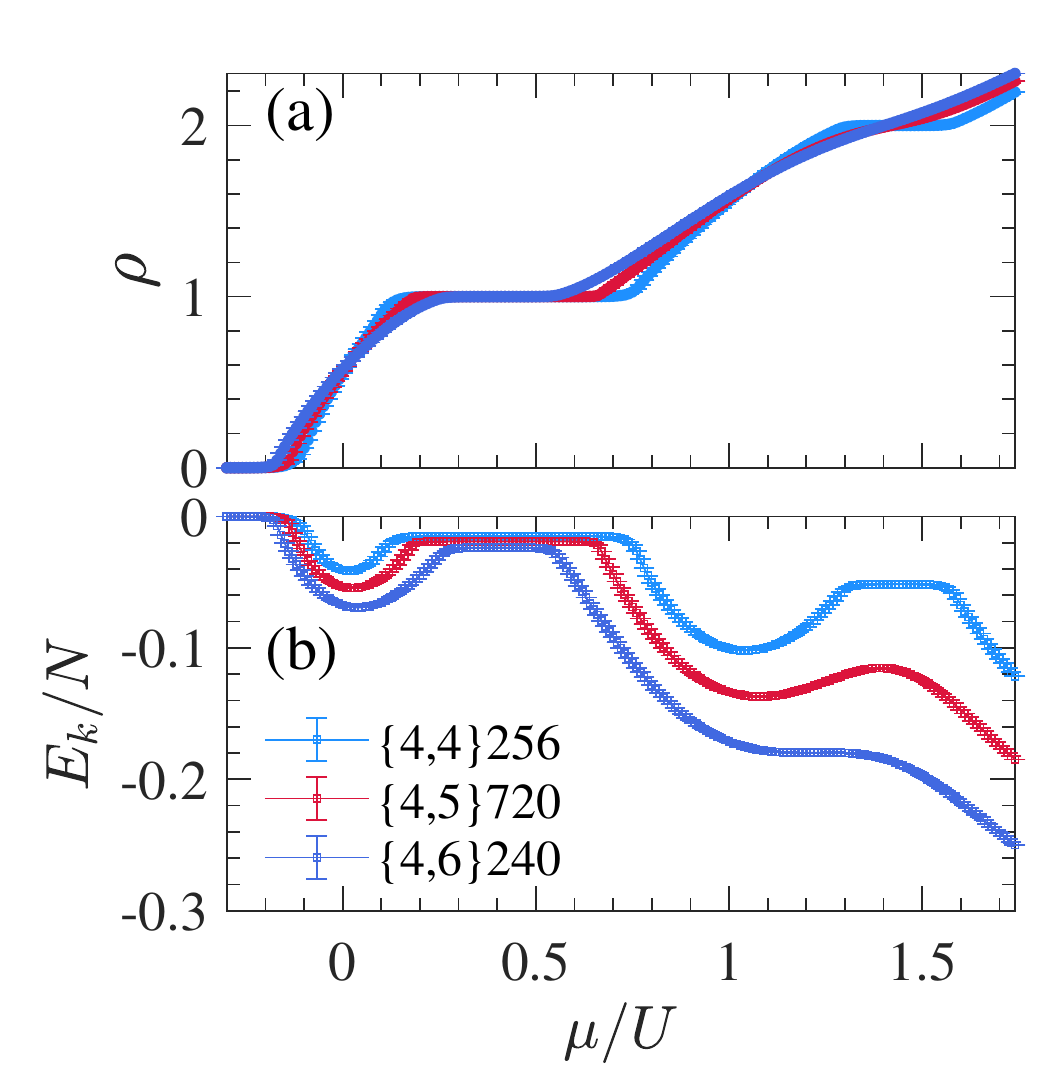} \caption{The average density (a) and the kinetic energy (b) as a function of the chemical potential for the normal square lattice \{4,4\}, and the hyperbolic lattices \{4,5\} and \{4,6\}. }
\label{fig4}
\end{figure}

\begin{figure*}[htbp]
\centering \includegraphics[width=14cm]{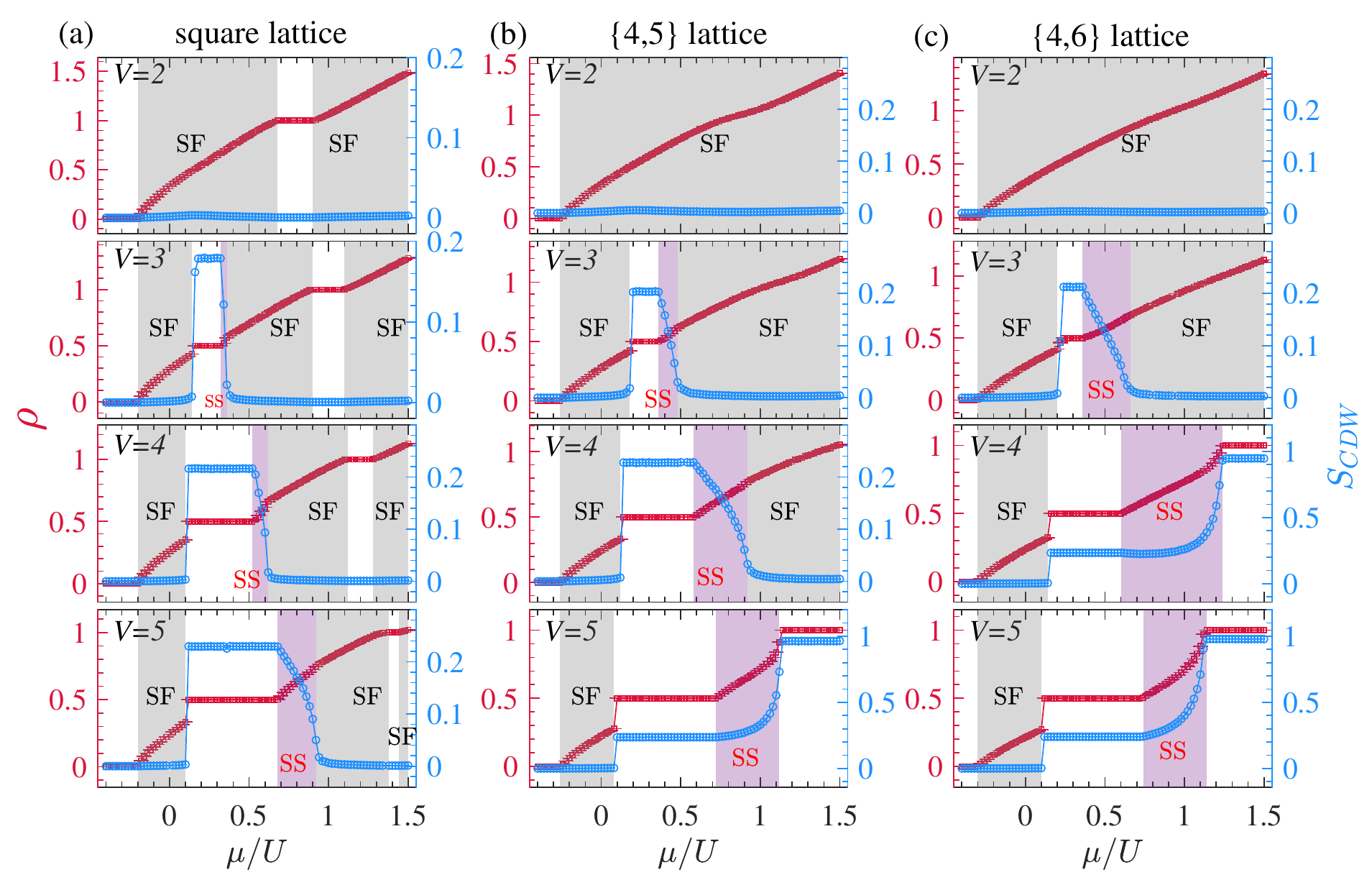} \caption{The average density and the static staggered structure factor as a function of $\mu$ at several values of $V$: (a) the square lattice; (b) $\{4,5\}$ and (c) $\{4,6\}$ hyperbolic lattices. Here $U=20t$ is used, and is taken as an energy scale of the chemical potential.}
\label{fig5}
\end{figure*}

The above phase diagrams are obtained by computing the compressibility $\kappa=\frac{\partial \rho}{\partial \mu}$ as a function of $\mu/U$ at constant $t/U$. Figure \ref{fig4} shows the average density $\rho$ as a function of $\mu/U$ on the cut with fixed
$t/U = 0.03$, along which the typical quantum phases of the phase diagram can be characterized. The average density $\rho$ exhibits a sequence of plateaus at integer fillings, on which $\kappa$ vanishes.
So the plateaus correspond to the incompressible Mott insulators, whose gaps are determined by the widths of the plateaus. Between
the insulators, the average density increases continuously
with the chemical potential and the compressibility has
a finite value, implying the system is in the superfluid phase.

We also calculate the average kinetic energy per site\cite{PhysRevB.80.014524},
\begin{align}\label{eq11}
E_{\bf k}=\frac{ -t\sum_{\langle i,j\rangle} \langle b_i^{\dagger}b_{j}+h.c.\rangle}{N},
\end{align}
as a function of $\mu/U$ for the same parameters as in Fig.\ref{fig4}(a). In the Mott insultor, the kinetic energy is greatly suppressed, and takes constant values, exhibiting a series of plateaus corresponding to those in the curve of the average density. When the system becomes superfluid, the kinetic energy is lowered, and takes a maximum value approximately at the medium density of two adjacent Mott insulators.

Next we include the NN repulsion $V$, which favors a staggered charge density wave (CDW). In order to characterize this phase, we calculate the static structure factor,
\begin{align}\label{eq12}
S_{ \textrm{CDW}}=\frac{1}{N}\sum_{ij}(-1)^{\textrm{sgn}(i,j)}\langle n_in_j\rangle,
\end{align}
where $\textrm{sgn}(i,j)=0(1)$ if $i,j$ belong to the same (opposite) sublattice. A perfect CDW with only occupied and unoccupied NN sites has $S_{ \textrm{CDW}}=N m^2/4$, where $m$ is the number of bosons on the occupied sites. Thus a CDW insulator is characterized by the plateaus of the static structure factor and the average density with the magnitudes $S_{ \textrm{CDW}}/N\sim m^2/4$ and $\rho=m/2 (m=1,2,...)$, respectively. The inclusion of the NN interaction will also generate an exotic supersolid phase, which is characterized by both nozero $S_{ \textrm{CDW}}$ and nozero $\kappa$.

Figure \ref{fig5} plots $\rho$ and $S_{ \textrm{CDW}}$ as a function of the chemical potential for several values of $V$ at $t/U=0.05$. As $V$ increases, there appears a $\rho=1/2$ plateau in the $\rho-\mu$ curve, and meanwhile the structure factor keeps constant with the value $S_{\textrm{CDW}}/N\sim 1/4$. Hence it is identified as a $\rho=1/2$ CDW insulator. Introducing holes to the CDW insulator makes the $\rho, S_{ \textrm{CDW}}$ curves discontinuous, implying the crystalline order is destroyed immediately by the holes. The instability is caused by the formation of domain walls, which leads to a phase separation between a $\rho=1/2$ insulator and a $\rho<1/2$ uniform superfluid. As shown in Fig.\ref{fig5}, such a behavior happens on both square and hyperbolic lattices.

It is well known that adding bosons to the $\rho=\frac{1}{2}$ CDW insulator can induces the supersolid phase on square lattice~\cite{PhysRevLett.94.207202}. Similarly, the supersolid phase can also be generated on hyperbolic lattices. At fixed $V$, the region of the supersolid phase is enlarged compared to that of the square lattice. Besides, the supersolid region expands as $q$ in the Schl\"afli symbol increases.

The above behavior can be understood qualitatively from the formation mechanism of the supersolid. The supersolid is most likely to happen with $zV\sim U$, when an added boson can be placed on either an occupied or unoccupied site since the total energy differs little for the two cases.
The boson can delocalize between the two sublattices to further lower the kinetic energy. The effective Hamiltonian in the two-state basis is approximated as follows
\begin{align}\label{eq13}
\tilde{ {\cal H} }=\left(
                     \begin{array}{cc}
                       U-\mu & zt \\
                       zt & zV-\mu \\
                     \end{array}
                   \right).
\end{align}
The total energy is directly obtained by diagonalizing the matrx, and we have the ground-state energy: $E=(U+zV)/2-\sqrt{ (zt)^2+\Delta^2 }-\mu$ with $\Delta=(U-zV)/2$. The kinetic energy is $-\sqrt{ (zt)^2+\Delta^2 }\propto t$ for small $\Delta$, which is large. Hence it is energetically favorable for the doped bosons to hop and form a superfluid on top of the CDW background, realizing a supersolid.

The coordination number of a hyperbolic lattice is $z=q$. Thus as $q$ increases, the supersolid can be stabilized by smaller $V(\sim U/z)$. For the $V=3t$ cases in Fig.\ref{fig5}, $zV$ with $z=6$ is closest to $U$ (here $U=20t$ is used), thus the supersolid region of the $\{4,6\}$ hyperbolic lattice is the largest.

\section{Conclusions}
We studied the Bose--Hubbard model on bipartite $\{ 4,5\}$ and $\{ 4,6\}$ hyperbolic lattices using SSE QMC simulations. In the presence of only on-site interaction, the phase diagram contains Mott insulators at integer fillings and incommensurate superfluid. As $q$ in the Schl\"afli symbol increases, the size of the Mott insulating lobes shrink. It is caused by the increase of the kinetic energy, which is proportional to the coordination number. By further including NN interaction, there appear staggered CDW at half integer fillings and exotic supersolid states. It is found that the supersolid is stabilized at smaller $V$ for larger $q$. We qualitatively analyze the underlying mechanism for this behavior. Our results suggest that the hyperbolic lattices provide a unique platform to study the effect of the coordination number on quantum phase transitions in Bose--Hubbard model. With the remarkable progress in cold-atom systems, the extended Bose--Hubbard models have been realized experimentally with extremely tuneability and cleaness~\cite{landig2016quantum}. Besides, the assembly of defect-free, arbitrarily shaped arrays of optical traps using holographic methods and fast, programmable moving tweezers has been reported~\cite{Barredo1021,barredo2018synthetic}. Thus it is very possible that our results will be experimentally realized in the setups with ultracold atoms in optical lattices.

\section{Acknowledgments}
H.G. acknowledges support from the NSFC grant
Nos. 11774019 and 12074022, the Fundamental Research Funds for the
Central Universities and the HPC resources at Beihang
University. X.Z. and S.F. are supported by the National
Key Research and Development Program of China under
Grant No. 2016YFA0300304, and NSFC under Grant
Nos. 11974051 and 11734002.
NPB acknowledges support through the UCLQ fellowship and the EPSRC Prosperity Partnership in Quantum Software for Simulation and Modelling (EP/S005021/1).


\appendix
\section{The second-order perturbation theory}
The analytical form of the self-consistent equation can be obtained using the second-order perturbation theory. In order to do this, we start from Eq.(\ref{eq4}), and have $H_i^{loc}|N_i\rangle=\epsilon_{N_i}|N_i\rangle$ with
\begin{align}
\epsilon_{N_i}=-\mu N_i +\frac{U}{2}N_i (N_i-1).
\end{align}
The ground state $|N_i^{loc}\rangle$ ($N_i^{loc}$ is a positive integer) is then obtained  by minimizing $\epsilon_{N_i}$, and we have:
\begin{equation}
\left\{
\begin{aligned}
&N_i^{loc}=0 \qquad\qquad\qquad\qquad \rm{if}\quad \mu \le 0\\
&N_i^{loc}-1< \frac{\mu}{U} \le N_i^{loc}\qquad\ \, \rm{if}\quad \mu > 0.
\end{aligned}
\right.
\end{equation}

Near the phase boundary, the value of $\Psi$ is small. Hence the term $M=-zt(\Psi^* b_i^{\dagger}+\Psi b_i)$ in Eq.(4) can be taken as a perturbation, and the ground-state energy is directly calculated using the second-order perturbation theory:
\begin{align}
&E_0^i=\epsilon_{N_i^{loc}}+a_2^i|\Psi|^2+O(|\Psi|^4), \\
&a_2^i|\Psi|^2=zt|\Psi|^2+\sum_{N_i\neq N_i^{loc}}\frac{|\langle N_i^{loc}|M|N_i\rangle|^2}{\epsilon_{N_i^{loc}}-\epsilon_{N_i}}. \nonumber
\end{align}

Using the relations
\begin{align}
\langle N_i^{loc} |M| N_i^{loc}+1 \rangle&=-zt\Psi\sqrt{N_i^{loc}+1}, \\
\langle N_i^{loc} |M| N_i^{loc}-1 \rangle&=-zt\Psi^* \sqrt{N_i^{loc}}, \nonumber
\end{align}
we obtain
\begin{align}
&a_2^i|\Psi|^2=zt|\Psi|^2+ \\
&z^2 t^2|\Psi|^2\left(\frac{N_i^{loc}+1}{\mu-UN_i^{loc}}+\frac{N_i^{loc}}{-\mu+U(N_i^{loc}-1)}\right). \nonumber
\end{align}

The phase boundary in the plane ($t/U,\mu/U$) is defined by the condition $a_2^i=0$, thus the following self-consistent equation is reached,
\begin{align}
\frac{1}{t/U}=-z\left(\frac{N_i^{loc}+1}{\mu/U-N_i^{loc}}+\frac{N_i^{loc}}{-\mu/U+(N_i^{loc}-1)}\right).
\end{align}

The above equation is solved numerically, and the obtained transition lines for $\{4,4\}$ square, $\{4,5\}$ and $\{4,6\}$ hyperbolic lattices are almost the same with those from the mean-field approximation.

\nocite{*}
\bibliography{hyperbolic_ref}

\end{document}